# Molecular Determinants of Orthosteric–allosteric Dual Inhibition of PfHT1 by Computational Assessment

Decheng Kong[1,*], Jinlong Ren[1,*], Zhuang Li[1], Guangcun Shan[2,3,5,6,†], Zhongjian Wang[2], Ruiqin Zhang[3], Wei Huang[5], Kunpeng Dou[1,4,†]

[1]College of Physics and Optoelectronic Engineering, Faculty of Information Science and Engineering, Ocean University of China, Qingdao 266100, China.

[2] BCPMdata Pharma Technology (Chengdu) Co. Ltd, Chengdu 610095, China

[3]Department of Physics, City University of Hong Kong, Hong Kong SAR 999077, China

[4]Engineering Research Center of Advanced Marine Physical Instruments and Equipment of Education Ministry of China & Key Laboratory of Optics and Optoelectronics of Qingdao, Ocean University of China, Qingdao 266100, China.

[5] School of Flexible Electronics, Sun Yat-sen University, Shenzhen 518107, China

[6] Institute of Precision Instrument and Quantum Sensing & Beijing Advanced Innovation Center for Big Data-based Precision Medicine, Beihang University, Beijing 100191, China.

*These authors contributed equally to this work.

†Corresponding authors

E-mail address: Guangcun Shan (gshan2-c@my.cityu.edu.hk), Kunpeng Dou (doukunpeng@ouc.edu.cn)

**Abstract**

To overcome antimalarial drug resistance, carbohydrate derivatives as selective PfHT1 inhibitor have been suggested in recent experimental work with orthosteric and allosteric dual binding pockets. Inspired by this promising therapeutic strategy, herein, molecular dynamics simulations are performed to investigate the molecular determinants of co-administration on orthosteric and allosteric inhibitors targeting PfHT1. Our binding free energy analysis capture the essential trend of inhibitor binding affinity to protein from published experimental IC$_{50}$ data in three sets of distinct characteristics. In particular, we rank the contribution of key residues as binding sites which categorized into three groups based on linker length, size of tail group, and sugar moiety of inhibitors. The pivotal roles of these key residues are further validated by mutant analysis where mutated to nonpolar alanine leading to reduced affinities to different degrees. The exception was fructose derivative, which exhibited a significant enhanced affinity to mutation on orthosteric sites due to strong changed binding poses. This study may provide useful information for optimized design of precision medicine to circumvent drug-resistant Plasmodium parasites with high efficacy.

**Keywords** PfHT1, Orthosteric–allosteric Dual-Pocket, Binding Free Energy, Molecular Dynamics, Chemical Backbone, Key Residues

# 1 Introduction

World Health Organization reported that about 263 million of malaria cases in 2023 alone [1]. The emergence of drug resistance against almost every front-line antimalarial drug is a major challenge. Novel mode of action is urgently needed to overcome drug resistance, for example, combinations of allosteric and orthosteric drugs [2], known as dualsteric modulators [3,4]. Hitherto, mounting efforts have been made to combat drug resistance by binding pharmacophores at both orthosteric and allosteric sites of targeted protein [5–8], such as case studies on mechanistic target of rapamycin (mTOR) [5], epidermal growth factor receptor (EGFR) [6,7], breakpoint cluster region-abelson1 (BCR-ABL1) kinase [8], androgen receptor (AR) [9].

To circumvent the artemisinin-resistant malaria parasites, Jiang et al. designed carbohydrate derivatives as inhibitors by dual targeting orthosteric and allosteric pockets of *P. falciparum* hexose transporter 1 PfHT1 [10,11]. Generally, two pharmacophores are needed to occupy the orthosteric site and the topologically distinct and distal allosteric site [2]. This recent elegant strategy benefited from achieving synergistic efficacy with a single chemical entity, simultaneously reaching both high potency and selectivity [10,11].

In addition, the pharmaceutical effect of this dual-inhibitor over PfHT1 varies depending on their structures [10,11], namely, sugar moiety, tail group, and a flexible linker $(CH_2)_n$, illustrating the importance of chemical backbone of the inhibitor. In contrast, compounds with very distinct structural backbones acting on human glucose transporters (hGLUTs) have been demonstrated in both exofacial inhibitors (phloretin and SA47) of GLUT3 [12] and endofacial inhibitors (Cytochalasin B and GLUT-i1) of GLUT1 [13].

Inspired by the experimental finding [10,11], we carried out molecular dynamics (MD) simulations in a follow-up study to further rank the molecular determinants of dual inhibition of PfHT1. To explore the structure-activity relationships, we classified 10 dual-inhibitors into three groups (Fig. 1 and table 1) with respect to the key features in experimental reference [10,11]. A systematic affinity profile of these inhibitors was created by integrating dynamics simulations, pocket volume analysis and energetics calculations. The simulations revealed that binding free energies showed a high correlation with experimental IC$_{50}$ data [10,11], validating the reliability of our analyses. We summarized and ranked the contributing of the key residues to the binding affinity in each of the three groups based on free energy decomposition. We further testified these observations using the mutation analysis. Glucose derivatives exhibited moderate and modest loss of binding affinity by mutation at allosteric or orthosteric sites. The outliers, fructose derivatives can be strongly sensitized by mutation at orthosteric sites due to changes of binding poses.

## 2 Methods and models

### 2.1 Calculation setup in molecular dynamics

Protein models were constructed by removing native ligands C3361 and glucose from PfHT1 complex crystal structures (PDB ID: 6M2L and 6M20 [10]). 6M20 was used as a template to complete the missing areas of 6M2L by Modeller [14]. PfHT1-inhibitor complex was then constructed using AutoDock Vina [15,16]. The optimized ligand-bound complex conformation from molecular docking was embedded into a 1-palmitoyl-2-oleoyl-sn-glycero-3-phosphocholine (POPC) lipid bilayer using the CHARMM-GUI Membrane Builder [17], solvated with TIP3P water [18] (22.5 Å thickness), and neutralized with 0.15 M NaCl. The dimensions of final system were 92 Å × 92 Å × 120 Å. Simulations were performed under physiological conditions (pH 7.0, 303 K) using the CHARMM36m force field [19] with WYF corrections [20] for the protein and lipids, CHARMM General Force Field [18] for ligands, respectively. Next, equilibration and production protocols were applied: (1) Energy minimization: 5,000 steps of steepest descent (SD) optimization to eliminate steric clashes. (2) Gradual restraint relaxation: Six-step equilibration under NVT/NPT ensembles with progressive reduction of positional restraints (initial values: 1,000 kJ·mol$^{-1}$·nm$^{-2}$ for lipids, 2,000 kJ·mol$^{-1}$·nm$^{-2}$ for protein–ligand; final values: 0/50 kJ·mol$^{-1}$·nm$^{-2}$ for lipids and protein–ligand). (3) Production MD simulation: 100 ns production simulation under NPT conditions to obtain thermodynamically stable conformational ensembles.

Here, the V-rescale temperature coupling was used to maintain the temperature of the membrane and protein at an interval of 1ps and the C-rescale was chosen for pressure coupling to maintain the system at one atmosphere at 5ps interval. The electrostatic calculation of the system was based on Particle-Mesh Ewald. The cut-off distance of non-bond interaction was 1.2 nm, and the switching distance was set to 1.0 nm. The LINCS algorithm was used to constrain the bonds containing hydrogen atoms to achieve a reasonable time step of 2 fs. Three parallel simulations were performed for all cases, the timescale of each was 100 ns. All MD calculations were based on GROMACS 2023.2 [21,22].

Based on the 6M20 system, steered molecular dynamics (SMD) simulations were initiated from the final frame of production MD simulation. TM1e helix was pulled towards the inhibitor molecule at constant-velocity (0.1nm/μs) by an umbrella potential in the NPT ensemble. The pull-groups were set to the Cα of Glu57 and the mass centers of TM10/TM12. In order to ensure the stability of the whole protein and avoid abnormal drift during the SMD simulations, the positional restraints of 100 kJ·mol$^{-1}$·nm$^{-2}$ was applied to the protein Cα atoms except TM1 helix. Next, the weighted histogram analysis method (WHAM) [23] in GROMACS was used to compute the potential of mean force (PMF) [24].

MM-GBSA calculations [25] are performed to obtain the binding free energy ($\Delta G_{binding}$) for evaluating molecular affinity [26].

## 2.2 Classification of atomic models for inhibitor

In this work, we classified the inhibitors into three types based on size of tail group, linker length, and sugar moiety as shown in Fig. 1c and table 1. The diversity of inhibitors in three components results in the structural complexity linked to their variances in biological activity.

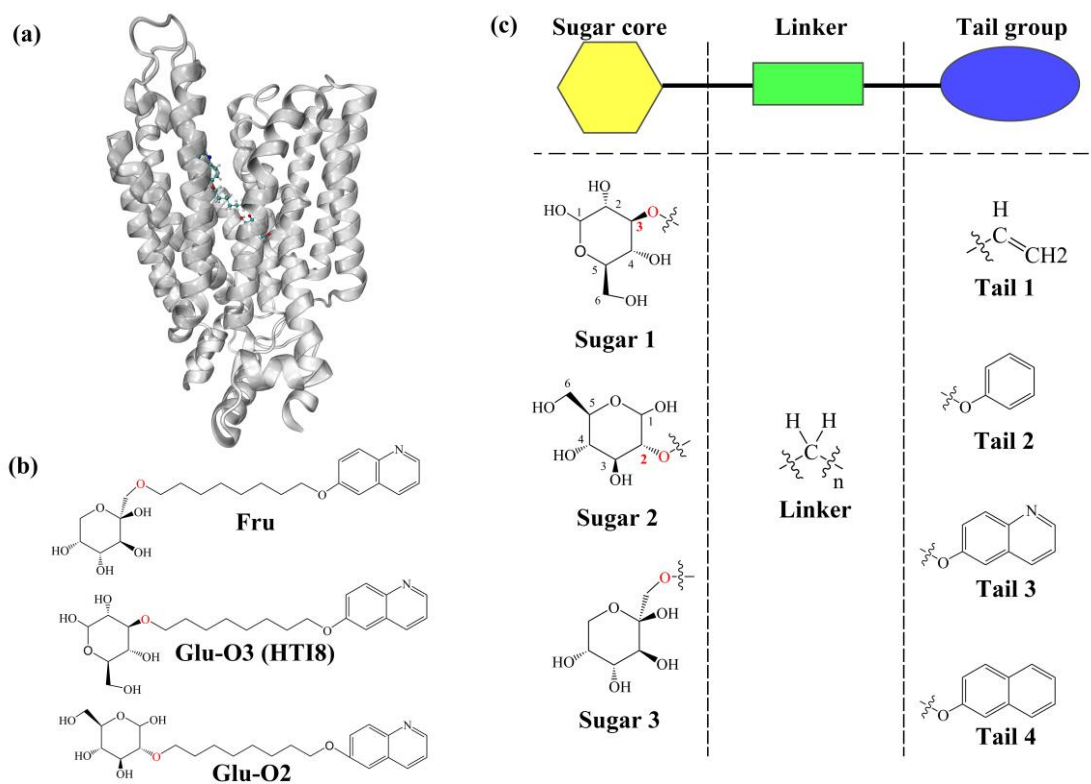

**Figure 1** (a) PfHT1-inhibitor complex. (b) Group III with variation of sugar moiety for evaluating orthosteric pocket. (c) The inhibitors features include three components, sugar core, aliphatic alkyl linker $(CH_2)_n$, and hydrophobic tail group.

Table 1 Inhibitors categorized into three groups based on linker length, size of tail group, and sugar moiety. The related atomic details are defined in Figs. 1 (b) and (c).

| Inhibitor name in present work and group classification | | | Experimental work[10,11] | |
|---|---|---|---|---|
| | | | Inhibitor name | IC$_{50}$ (μM) |
| Group I for evaluating allosteric pocket: Variation of linker (CH$_2$)$_n$ length; quinine as tail group; glucose as sugar core | n=6 | HTI6 | 2a[10] | 13.7 |
| | n=8 | HTI8 | 1c[10] | 0.513 |
| | n=9 | HTI9 | THPF-02[11] | 0.329 |
| | n=10 | HTI10 | 2b[10] | 0.473 |
| | n=12 | HTI12 | 2c[10] | 4.16 |
| Group II for evaluating allosteric pocket: Variation of tail group size; (CH$_2$)$_{n=8}$; glucose as sugar core | Tail1 | C3661 | C3661[10] | 33.1 |
| | Tail2 | HTI8-half | 1a[10] | 15.5 |
| | Tail4 | HTI8-N | 1b[10] | 2.37 |
| | Tail3 | HTI8 | 1c[10] | 0.513 |
| Group III for evaluating orthosteric pocket: Variation of sugar moiety; (CH$_2$)$_{n=8}$; quinine as tail group | linker (CH$_2$)$_8$ connected with glucose by C3 or C2 position | Glu-O3 (HTI8) | 1c[10] | 0.513 |
| | | Glu-O2 | THPF-03[11] | 1.22 |
| | linker (CH$_2$)$_8$ connected with fructose | Fru | 3b[10] | 128 |

## 3 Results and Discussion

MD simulations and post analysis are performed to obtain a detailed molecular description of inhibitor-target interaction.

### 3.1 Key residues for the formation and stabilization of allosteric pocket in PfHT1

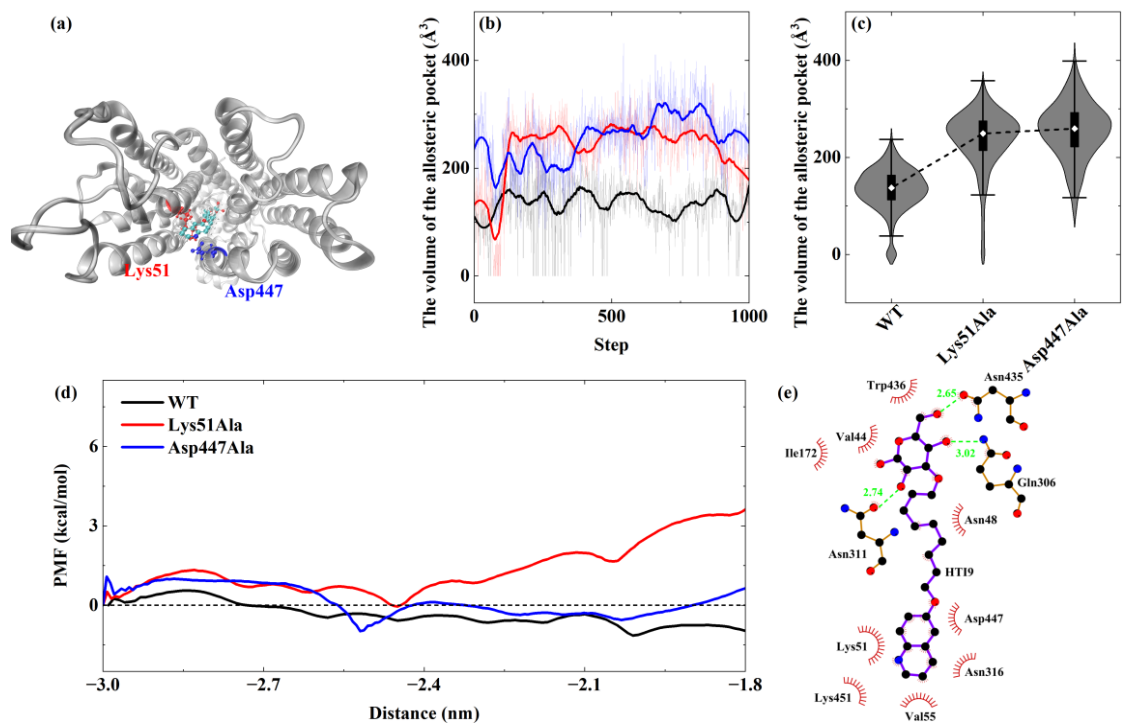

**Figure 2** (a) PfHT1-HTI9 complex. (b) and (c) Evolution of allosteric pocket volume for Lys51Ala, Asp447Ala mutation and WT counterpart. (d) Potential of the mean force (PMF) profile on formation and stabilization of the allosteric binding pockets for Lys51Ala, Asp447Ala mutation and WT counterpart. (e) Detailed interaction pattern of PfHT1-HTI9 complex.

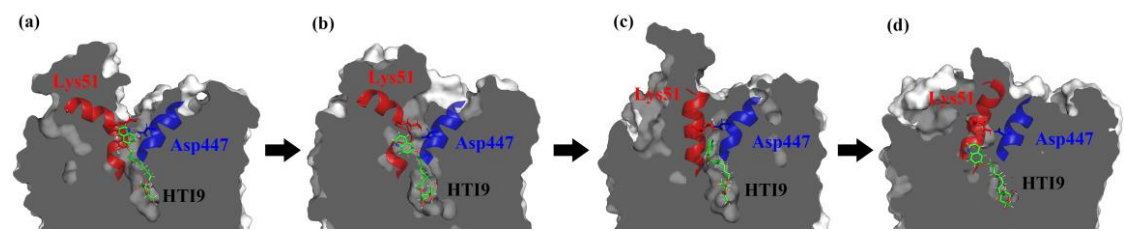

**Figure 3** MD snapshot of HTI9-PfHT complex in the wild-type during the closure of allosteric pocket. Dynamics behavior involved in the intermolecular interaction between Lys51 and the tail of HTI9 at the molecular level.

We take HTI9 (belongs to group II in table 1) as a model system to explore the molecular mechanism underlying the dynamic process of allosteric pocket formation based on static structural information from experiment work [10,11]. Aims at mimicking classical allosteric communication [27], translocation of HTI9 through the tunnel is expected to trigger the formation of allosteric pocket before its sugar core reaches the orthosteric site. However, after extensive MD simulations, no allosteric pocket is found to form after HTI9 arrives at the orthosteric site. Thus, we further probe the dynamic molecular basis of allosteric activation by mimicking reversed allosteric communication [28,29], namely, the perturbations from ligand binding at orthosteric site facilitating the closure of allosteric pocket. In this way, we succeed to probe the allosteric pocket shown in Fig. 2. We extracted the representative structure of different conformational states and identified the key residue Lys51 related to the formation of allosteric pocket whereas interaction between Lys51 and Asp447 correlated to the stabilization of the allosteric pocket (Fig. 3). This is in line with the experimental findings [10,11]. PMF plot was utilized to evaluate the formation barrier and consequent stability of allosteric binding pocket (Fig. 2(d)). In general, wild-type system displayed overall lower PMF values than mutant ones, implying good efficacy for more competent allosteric binding pocket. Higher PMF values and larger pocket volumes (Fig. 2(c)) in mutation systems indicated the impaired direct interaction between the residue Lys 51 and HTI9 by Lys51Ala whereas Asp447Ala leading to the destabilization (opening and loosening) of the pocket.

**3.2 Ranking the key residues for evaluating inhibitor affinity with orthosteric–allosteric dual-pocket in PfHT1**

**3.2.1 Linker length and allosteric pocket**

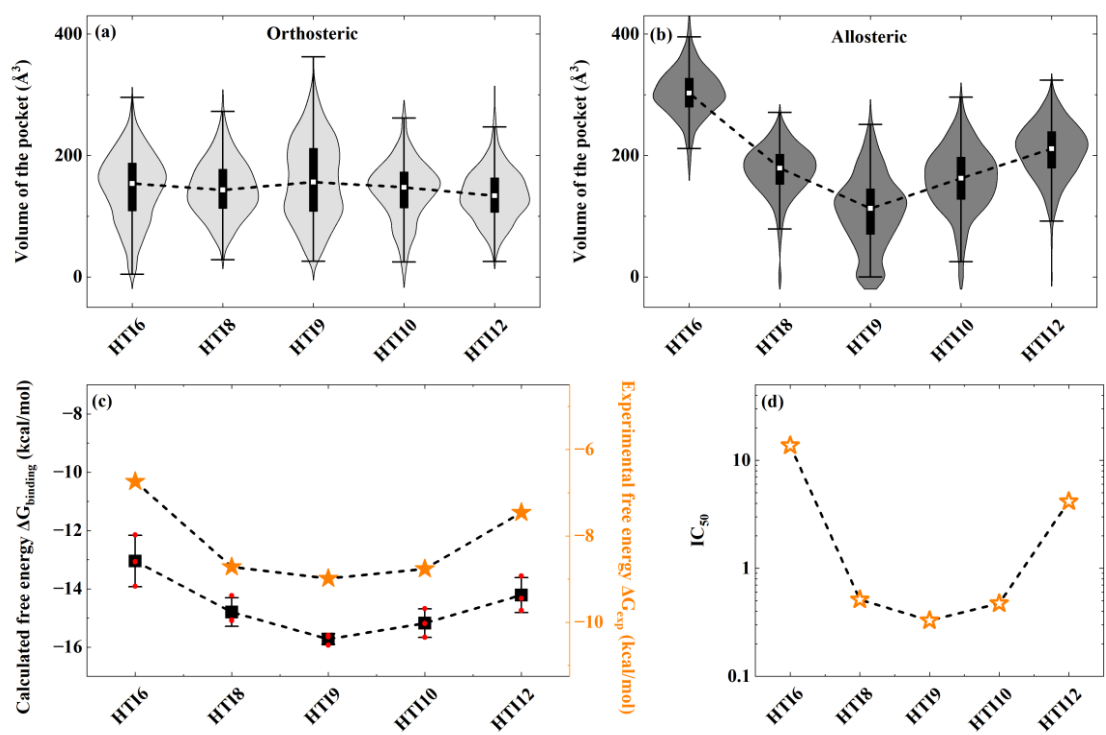

**Figure 4** Comparison of pocket volume for group I at (a) orthosteric site (b) allosteric site. (c) Comparison of binding free energy for calculated $\Delta G_{binding}$ and experimental $\Delta G_{exp}$= RTlnIC$_{50}$ for group I. (d) Experimental IC$_{50}$ trend for group I.[10,11]

The compounds in group I shared the same sugar core (glucose) and tail (quinine), with varied linker (CH$_2$)$_n$ length for evaluating allosteric binding, which engaged less well-conserved regulatory motifs outside the orthosteric pocket. Thus, they displayed similar orthosteric pocket volume but varied allosteric ones (Figs. 4(a) and (b)). The calculated binding free energy $\Delta G_{binding}$ exhibited nonmonotonic relationship with various linker length (Fig. 4(c)). This result was consistent with the in vitro experiments (Fig. 4(d)) [10,11]. The per-residue free energy decomposition (Figs. 5 and 6) unraveled that the contributions from Lys51, Val443 and Leu47 are the key determinants for binding affinity for this group. The binding affinity sum from these three residues (Fig. S1(d)) dominated the binding events at allosteric site which allowed repeating the experimental trend (Fig. 4(d)).

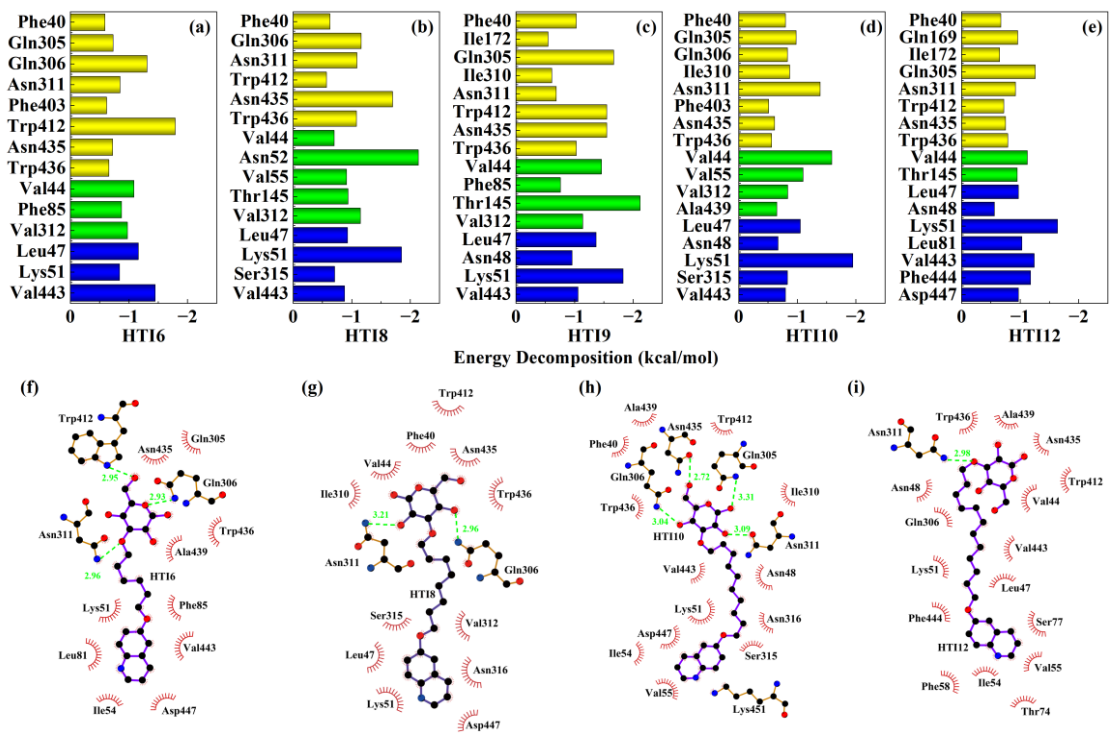

**Figure 5** Decomposed contribution of key residues to ΔG$_{binding}$ for group I (a)-(e). Yellow and blue related with orthosteric site and allosteric site, respectively. Binding interactions of inhibitor (f) HTI6, (g) HTI8, (h) HTI10 and (i) HTI12 to PfHT1.

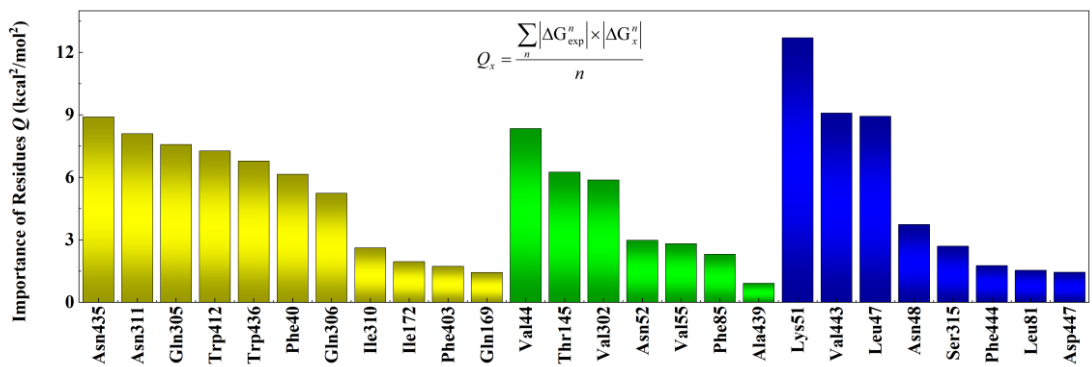

**Figure 6** Contributing assignment ranking of key residues to ΔG$_{binding}$ for group I.

### 3.2.2 Size of tail group and allosteric pocket

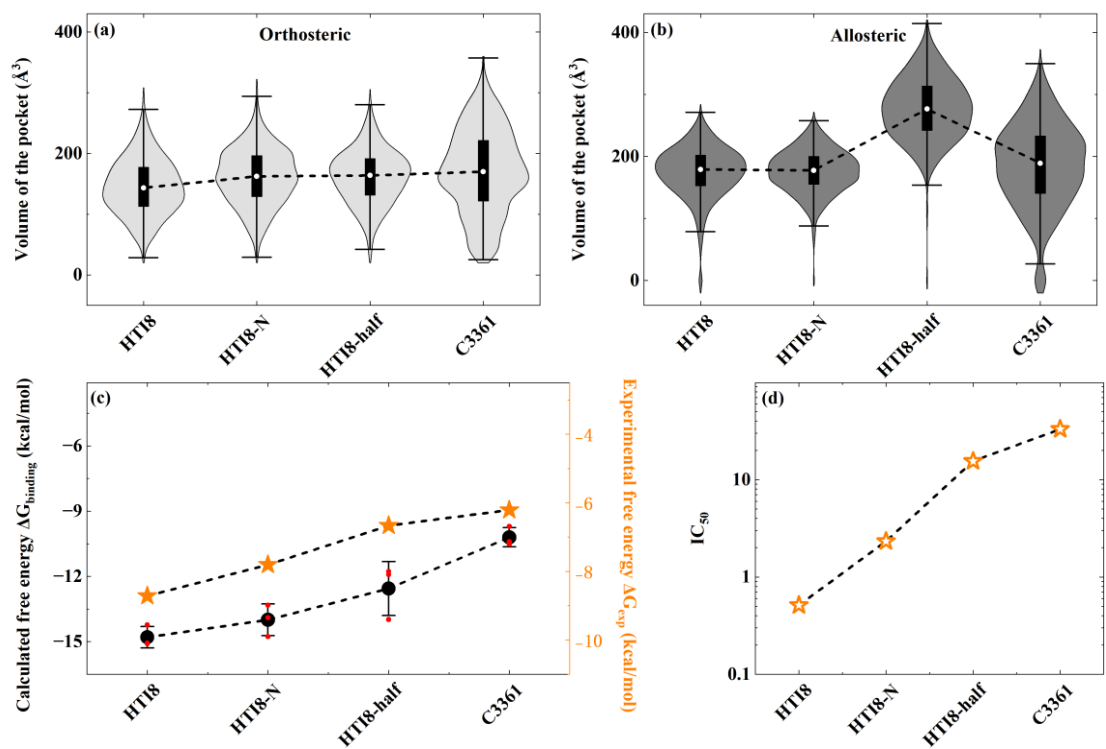

**Figure 7** Comparison of pocket volume for group II at (a) orthosteric site (b) allosteric site. (c) Comparison of binding free energy for calculated $\Delta G_{binding}$ and experimental $\Delta G_{exp}= RTlnIC_{50}$ for group II. (d) Experimental $IC_{50}$ trend for group II.[10,11]

This group explored the tail size for evaluating allosteric binding. The varied allosteric pocket volume did not capture the trend of experimental $IC_{50}$ trend (Figs. 7(b) and (d)). The exception of C3661 is contrary to the common scenario where enhanced pocket closure strengthens the binding affinity.

As shown in Fig. 7(c), energetic analysis still showed good agreement with experimental reference ($IC_{50}$ in Fig. 7(d)). Per-residue energy contribution (Figs. 8, 9 and S2) revealed that the hydrophobic bonding interactions between inhibitor and the critical residues Lys51, Leu47 and Val443 were responsible for the binding affinity in this group. The small allosteric pocket of C3661 did not imply strong affinity due to the reduced hydrophobic binding between Lys51, Leu47 and C3661 (Fig. 8 (d)).

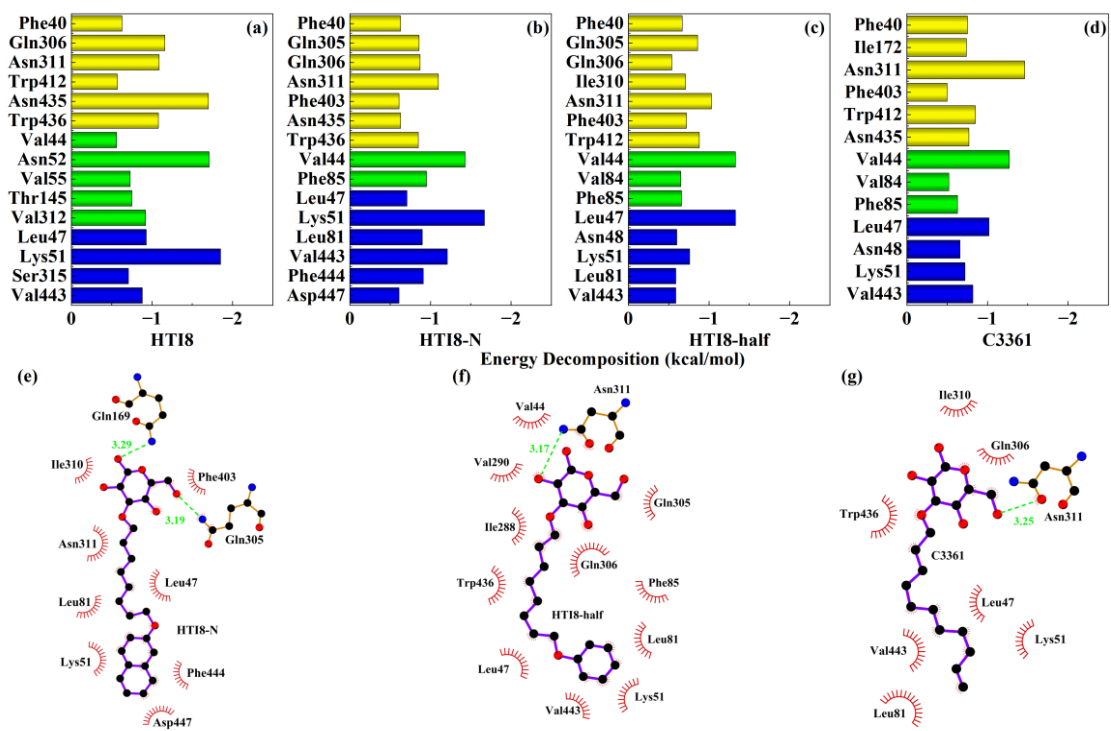

**Figure 8** Decomposed contribution of key residues to $\Delta G_{binding}$ for group II (a)-(d). Yellow and blue related with orthosteric site and allosteric site, respectively. Binding interactions of inhibitor (e) HTI8-N, (f) HTI8-half and (g) C3661 to PfHT1.

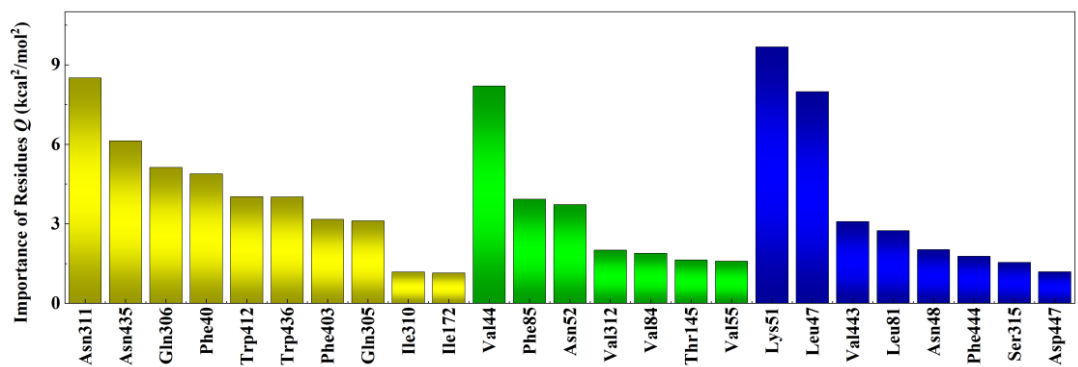

**Figure 9** Contributing assignment ranking of key residues to $\Delta G_{binding}$ for group II.

### 3.2.3 sugar moiety and orthosteric pocket

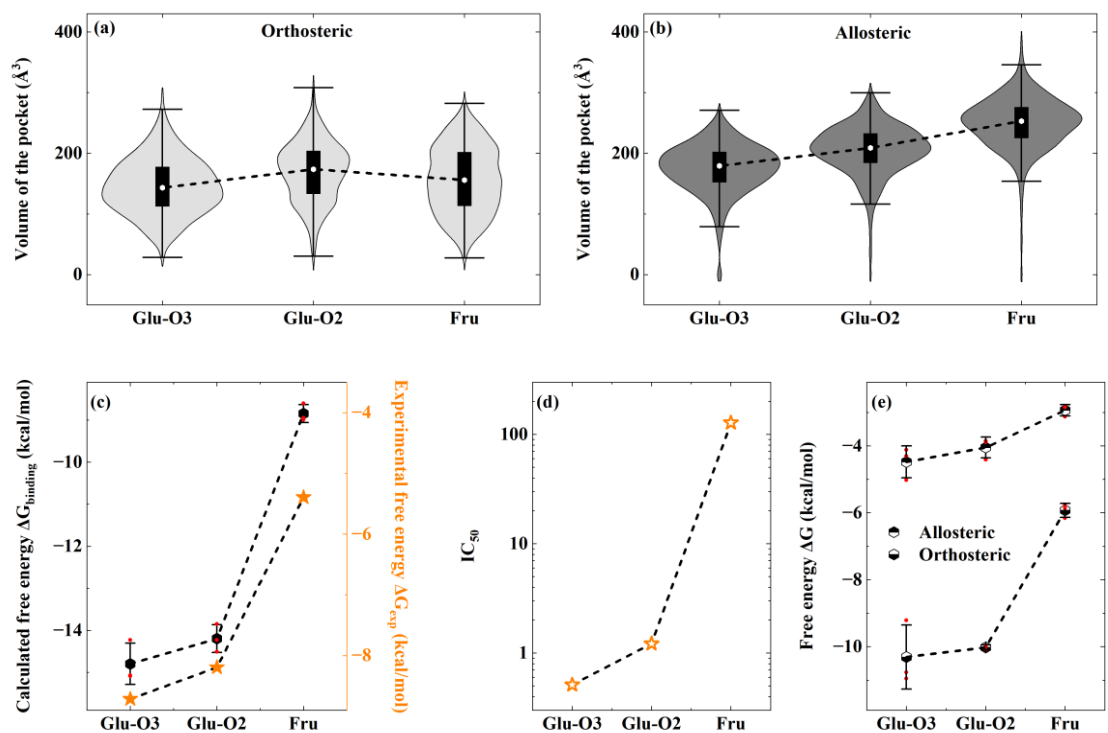

**Figure 10** Comparison of pocket volume for group III at (a) orthosteric site (b) allosteric site. (c) Comparison of binding free energy for calculated $\Delta G_{binding}$ and experimental $\Delta G_{exp}= RT\ln IC_{50}$ for group III. (d) Experimental $IC_{50}$ trend for group III.[10,11] (e) Decomposed binding free energy at orthosteric and allosteric site, respectively.

**Table 2** Free energy decomposition for group III at orthosteric site

| Residues | Free energy decomposition for group III (kcal/mol) | | |
|---|---|---|---|
| | Glu-O3 (HTI8) | Glu-O2 | Fru |
| Phe40 | -0.80 | -0.56 | -0.79 |
| Gln305 | -0.50 | -0.46 | -0.50 |
| Gln306 | -0.60 | -0.68 | -0.58 |
| Ile310 | -0.44 | | |
| Asn311 | -1.29 | -0.95 | -1.42 |
| Trp390 | -0.19 | -0.90 | -0.92 |
| Phe403 | -0.21 | | |
| Asn435 | -0.90 | -1.09 | -0.62 |
| Trp436 | -0.60 | -1.09 | |

This group aimed at evaluating orthosteric binding. As shown in Fig. 10(c), the glucose derivatives Glu-O3 and Glu-O2 showed an obviously higher binding potency than fructose one, Fru. The moderate and modest changes in dual pockets of this group could not account for this binding behavior (Figs. 10(a) and (b)). Free energy decomposition suggested that binding affinity at orthosteric site rather than at allosteric one played the dominant role to capture the experimental trend (Figs. 10 (d) and (e)). This reflected the binding specificity of sugar transport in PfHT1 [30]. It is noted that the identified key residues for binding carbohydrate derivatives in table 2 partial mismatched the essential ones (Q169, Q305, Q306, N311 and N341) at orthosteric site for uptaking of glucose and fructose by PfHT1 [30].

### 3.3 Validation of the interactions between key residues of PfHT1 and inhibitor by single point mutation

#### 3.3.1 mutation at allosteric sites

We further testified the key residues for binding affinity using the mutation analysis. For group I and II focusing on allosteric sites, binding free energy calculations revealed that Lys51 contribute in a significant way to modulate the binding affinity (Fig. 11). As unveiled in section 3.1, Lys51 not only provided the internal interactions with the tail of inhibitor, but also the coupling with Asp447 to stabilize the allosteric pocket. It is unraveled that most mutations have only a moderate or modest effect on C3661 binding, whereas Asp447Ala can cause strong deactivation to this inhibitor (Fig. 11(b)).

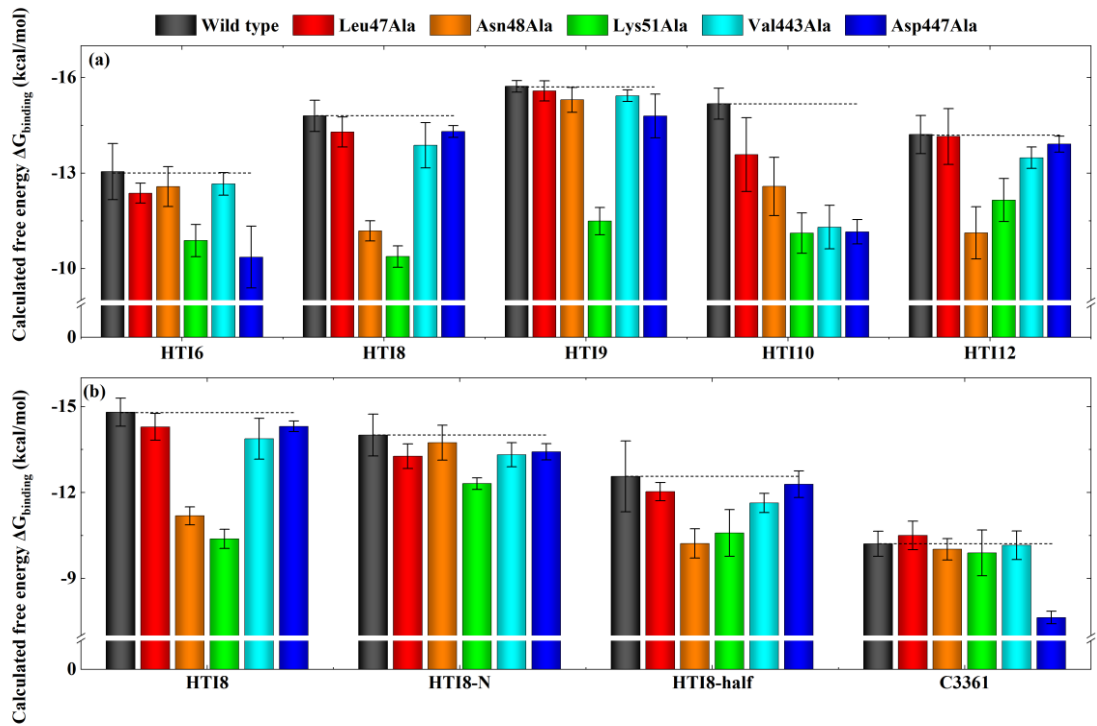

**Figure 11** Comparison of binding free energy $\Delta G_{binding}$ of wild type, Leu47Ala, Asn48Ala, Lys51Ala, Val443Ala, Asp447Ala for (a) group I (b) group II.

### 3.3.2 mutation at orthosteric sites

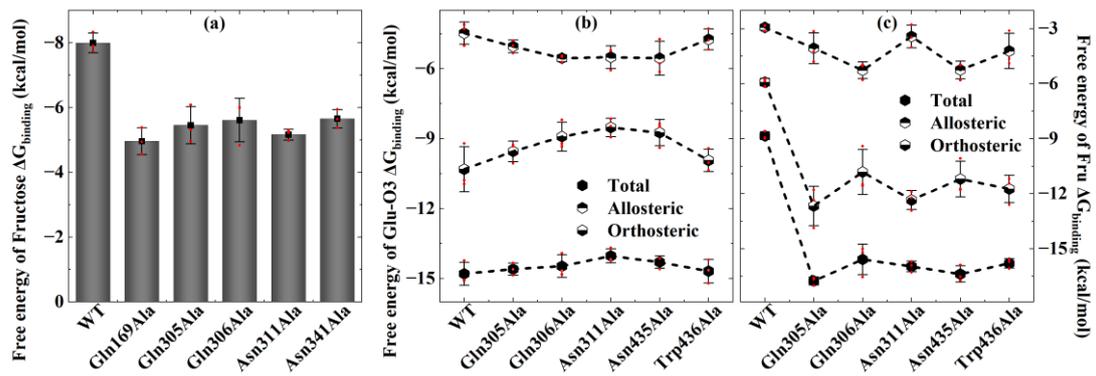

**Figure 12** Binding free energy $\Delta G_{binding}$ of fructose in wild type and mutant systems. Comparison of $\Delta G_{binding}$ for (b) Glu-O3 (c) Fru in wild type and mutant systems.

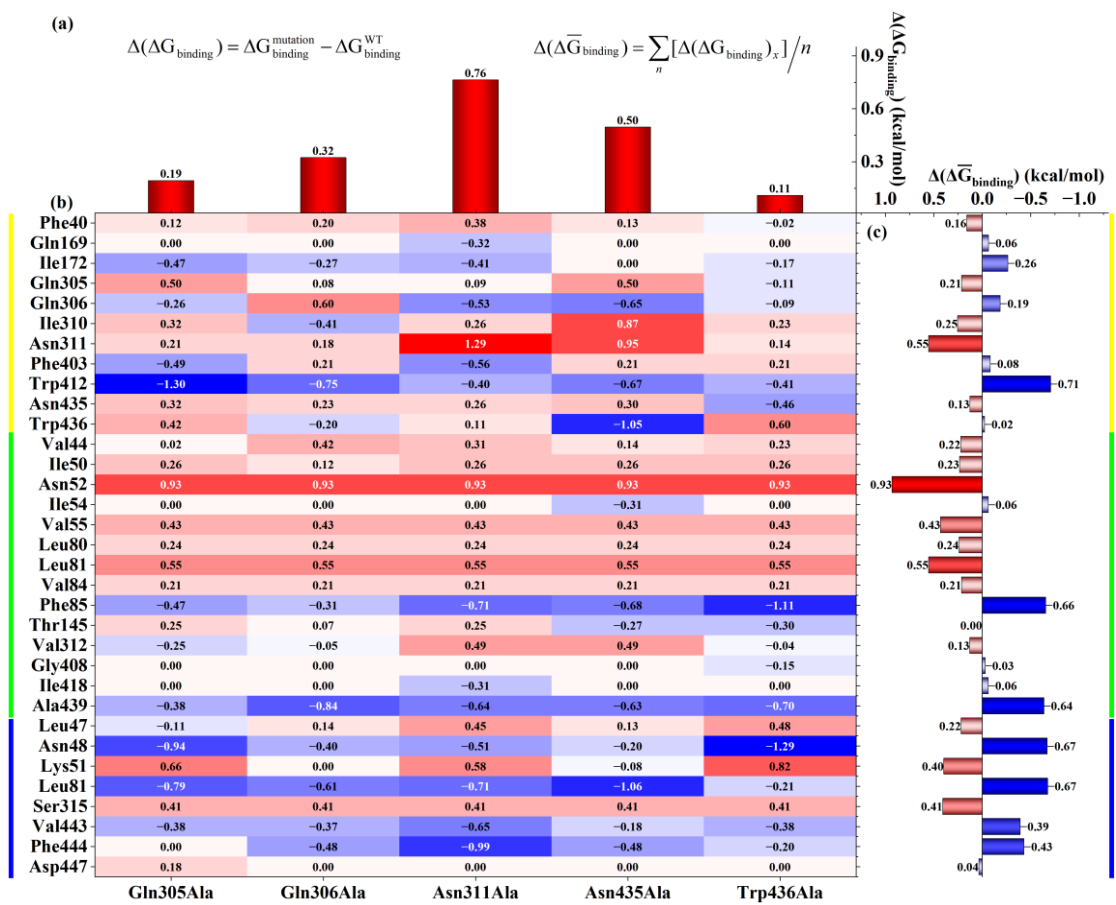

**Figure 13** Relative binding free energies for Glu-O3 in mutant systems with respect to wild type. Positive/negative values indicate that weaker/stronger interaction between inhibitor and the residue after mutation. Yellow and blue lines at both sides mark the residues at orthosteric site and allosteric site, respectively.

Mutations at orthosteric site have a minor effect on the binding affinity of Glu-O3 with glucose as sugar core (Fig. 12 (b)). As depicted in Fig. 13(a), mutations generally disrupted the interaction between the key residues in the orthosteric pocket and Glu-O3. In contrast, mutations located at orthosteric site caused strong sensitization to fructose derivative, Fru, as shown Fig. 12(c). Previous work indicated mutations at conserved residues of orthosteric pocket impaired transport of fructose by PfHT1[30]. Our binding free energy $\Delta G_{binding}$ of fructose in wild type and mutant systems also verified this trend (Fig. 12(a)). To elucidate the outlier mutation effect in PfHT1-Fru complex, free energy decomposition was performed and suggested that other residues emerged as new binding sites (Fig. 14 (c)). This is due to mutation will greatly change the binding pose of Fru (Fig. 15(b)) whereas marginal effect on that of Glu-O3 (Fig. 15(a)).

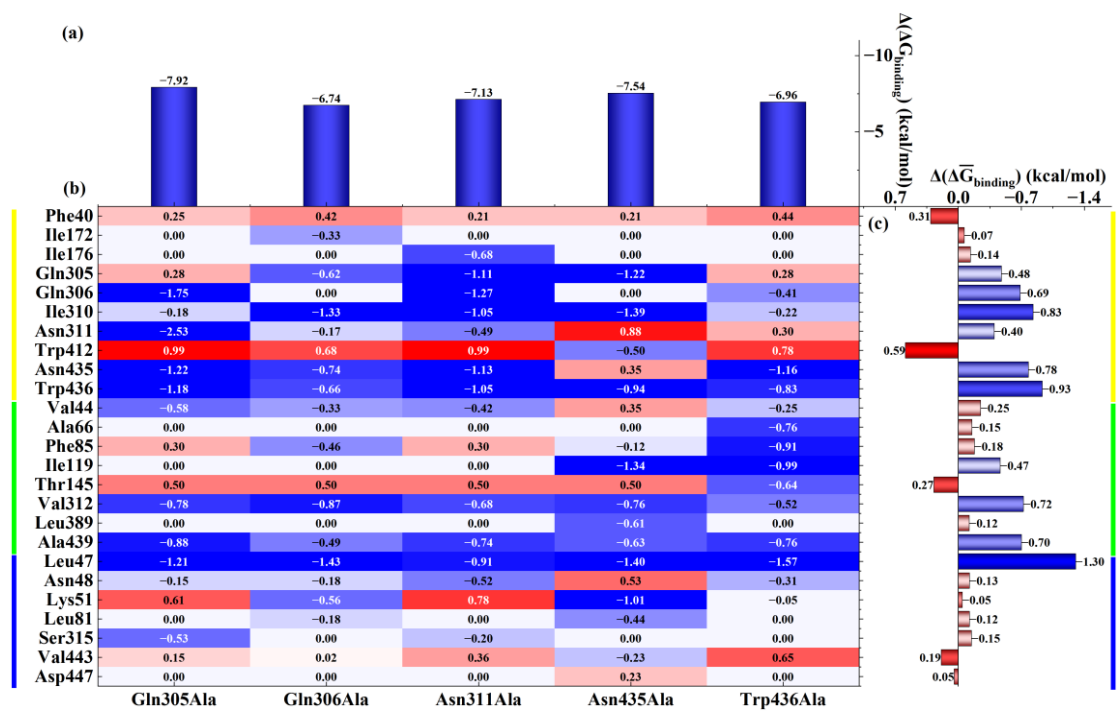

**Figure 14** Relative binding free energies for Fru in mutant systems with respect to wild type. Positive/negative values indicate that weaker/stronger interaction between inhibitor and the residue after mutation. Yellow and blue lines at both sides mark the residues at orthosteric site and allosteric site, respectively.

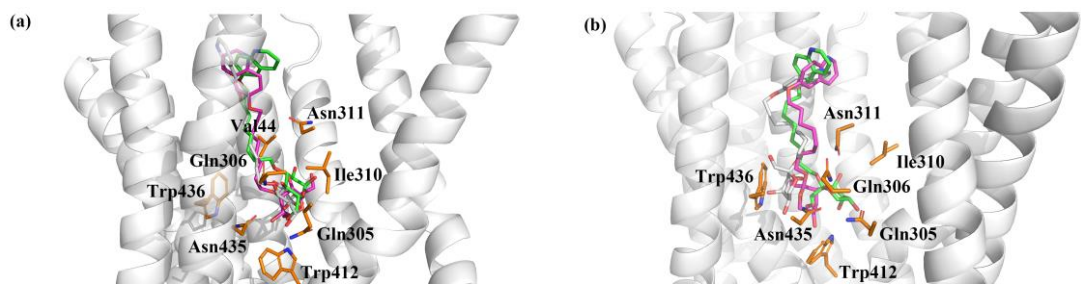

**Figure 15** Cartoon representation of PfHT1 with (a) Glu-O2 and (b) Fru. The interacting residues are labelled by orange. Structural superimposition of colored by green, and pink, silver from wild-type, mutant complexes, respectively.

## 4 Conclusions

Validated by the strong agreement between our calculated protein−inhibitor affinities (based on binding free energy) and experimental references ($IC_{50}$ values), we rank the key molecular determinants for orthosteric–allosteric dual inhibition of PfHT1 based on systemic case studies covering the three essential building blocks of inhibitors (linker length, size of tail group, and sugar moiety). Mutation studies further revealed that targeted perturbations at the orthosteric site can significantly enhance the inhibitory potency of fructose-derived compounds, likely due to its strong conformational adjustments in binding pose. Notably, this outlier case underscores the nuanced role of subatomic interactions and conformational dynamics in fine-tuning inhibitor efficacy for the structure-based drug design. In summary, these insights here advance the mechanistic understanding of resistance-countering drug design, emphasizing the need to integrate molecular flexibility and residue-specific interactions into strategies targeting artemisinin-resistant malaria parasites.

**Declarations**

The authors declare that there is no conflict of interest.

**Electronic supplementary materials**

Supplementary data to this article can be found online.

**Acknowledgments**

The work was carried out at Marine Big Data Center of Institute for Advanced Ocean Study of Ocean University of China and supported by Fundamental Research Funds for the Central Universities (Grant no.3002000−842364006), Key R&D Program of Shandong Province (Grant no.2023CXPT101), Zhoushan Oceanthink Marine Science & Technology Co., Ltd. (Grant no.3002000−961236054100). We thank Dr. Chen at Institute of Automation, Chinese Academy of Sciences and Prof. Juan de Pablo at New York University for providing technical support and helpful discussions.